\documentclass[aps,preprint]{revtex4}%
\usepackage{amsfonts}
\usepackage{amsmath}
\usepackage{amssymb}
\usepackage{graphicx}%
\providecommand{\U}[1]{\protect\rule{.1in}{.1in}}
\begin{document}
\preprint{ }
\title[ ]{Charge dynamics and "in plane" magnetic field I: Rashba-Dresselhauss
interaction, Majorana fermions and Aharonov-Casher theorems}
\author{Diego Julio Cirilo-Lombardo}
\affiliation{Instituto de Fisica del Plasma (INFIP), Consejo Nacional de Investigaciones
Cientificas y Tecnicas (CONICET), Buenos Aires, Argentina }
\affiliation{Bogoliubov Laboratory of Theoretical Physics, Joint Institute for Nuclear
Research, 141980, Dubna (Moscow region), Russian Federation}
\author{}
\affiliation{}
\author{}
\affiliation{}
\pacs{PACS number}

\begin{abstract}
The 2-dimensional charge transport with parallel (in plane) magnetic field is
considered from the physical and mathematical point of view. To this end, we
start with the magnetic field parallel to the plane of charge transport, in
sharp contrast to the configuration described by the theorems of Aharonov and
Casher where the magnetic field is perpendicular. We explicitly show that the
specific form of the arising equation enforce the respective field solution to
fulfil the Majorana condition. Consequently, as soon any physical system is
represented by this equation, the rise of fields with Majorana type behaviour
is immediately explained and predicted. In addition, there exists a quantized
particular phase that removes the action of the vector potential producing
interesting effects. Such new effects are able to explain due the geometrical
framework introduced, several phenomenological results recently obtained in
the area of spintronics and quantum electronic devices. The quantum ring as
spin filter is worked out in this framework and also the case of the quantum
Hall effect.

\end{abstract}
\volumeyear{year}
\volumenumber{number}
\issuenumber{number}
\eid{identifier}
\date[Date text]{date}
\received[Received text]{date}

\revised[Revised text]{date}

\accepted[Accepted text]{date}

\published[Published text]{date}

\startpage{101}
\endpage{102}
\maketitle
\tableofcontents

\section{Introduction}

In 1937 Ettore Majorana propose a new representation to the celebrated Dirac
equation, where the components of the spinor solution are related themselves
by complex conjugation [11]. Due his personal problems, he could not have
foreseen the whirlwind of activity that would follow: not only in particle
physics but also in nanoscience and condensed matter physics (for a review
about this issue in condensed matter see [14]). The recent storm of activity
in condensed matter physics has focused on the `Majorana zero modes' i.e.
emergent Majorana-like states occurring at exactly zero energy that have a
remarkable property of, if they are considered as particles, being their own
antiparticles (self-conjugated). Sometimes, this property is expressed as an
equality between the particle's creation and annihilation operators. As we
will see below, there exists the\ general idea that any ordinary fermion can
be though of as composed of two Majorana fermions: this is only a partial
picture, the real fact is that there exists a particular representation were a
fermion efectively can be represented as bilinear combination of two states of
fractionary spin [8]. For example, the considered in condensed matter physics
"Majorana zero modes" are believed to exhibit the so called non-Abelian
exchange statistics [4, 5] which endows them with a technological potential as
building blocks of future quantum memory immune against many sources of
decoherence. Recent advances in our understanding of solids with strong
spin-orbit coupling, combined with the progress in nanofabrication, put the
physical realization of the Majorana states to be considered as possible. In
fact signatures consistent with their existence in quantum wires coupled to
conventional superconductors and other type of devices have been reported by
several groups [12].

By the other hand and with other motivations, Aharonov and Casher [1] proved
two theorems for the case of a 2-D magnetic field. The first theorem states
that an electron moving in a plane under the influence of a perpendicular
inhomogeneous magnetic field has N ground-energy states, where N is the
integral part of the total flux in units of the flux quantum $\Phi_{0}%
=2\pi/e\equiv hc/e$ (m=1)$.$The corresponding Dirac equation for the
Aharonov-Casher-Theorem (ACT)\ configuration is\footnote{We denote the fixed
reference system as X,Y,Z and the coordinates in plane by $x_{1},$ $x_{2},$
$x_{3}$.}
\begin{equation}
\left[  \sigma_{x}\left(  \partial_{x}-ieA_{x}\right)  +\sigma_{y}\left(
\partial_{y}-ieA_{y}\right)  \right]  \varphi=0 \tag{A}%
\end{equation}
We introduce the transformation%
\begin{equation}
\psi=e^{e\phi\sigma_{z}}\varphi\tag{B}%
\end{equation}
this transformation (phase) permits us to eliminate the magnetic field
explicitly from the Dirac equation where $\phi$ satisfies the relations%
\[
\partial_{x}\phi=A_{y},\text{ \ \ \ \ \ \ }\partial_{y}\phi=-A_{x}%
\]
and $\varphi$ is eigenfunction of $\sigma_{z}$ $\left(  \sigma_{z}\varphi
_{s}=s\varphi_{s}\right)  $. Having into account that $B\left(  x,y\right)
=\partial_{x}A_{y}-\partial_{y}A_{x}$ we arrive to
\[
B\left(  x,y\right)  =\left(  \frac{\partial^{2}}{\partial x^{2}}%
+\frac{\partial^{2}}{\partial y^{2}}\right)  \phi
\]
It is easy to see that (asymptotically) for $r\rightarrow\infty$ we have%
\[
\phi\left(  x,y\right)  =\frac{\Phi}{2\pi}\ln\left(  \frac{r}{r_{0}}\right)
\]
where%
\[
\Phi=\int B\left(  x,y\right)  dxdy
\]
is the total magnetic flux through the $\left(  x,y\right)  $-plane, $r_{0}$
is some real constant. Consequently we immediately obtain%
\[
\varphi_{s}=\left(  \frac{r_{0}}{r}\right)  ^{\frac{\Phi s}{\Phi_{0}}}\psi
_{s}\left(  w\right)
\]
where $w=x+isy$ and $\psi_{s}\left(  w\right)  $ is an entire function of
$w$\ because after the elimination of the magnetic field from the equation
$\left(  A\right)  $ it takes the simplest form%
\[
\left(  \partial_{x}+is\partial_{y}\right)  \psi_{s}\left(  w\right)  =0
\]
In order that $\varphi_{s}$ to be square integrable function we should
consider $\Phi s>0$ and $\psi_{s}$ has to be a polynomial whose degree is not
greater than $N-1$, where $N=\left\{  \Phi/\Phi_{0}\right\}  $, obtaining $N$
independient solutions for $\psi_{s}:$ $1,w,w^{2},....,w^{N-1}.$Through this
paper the same procedure as for the ACT\ configuration will be performed but
in the case of "in plane" (parallel) magnetic field [2,3,4,5].

The plan of this paper is as follows. Section II we obtain the conditions
whether the magnetic field parallel to the charge transport can be "removed"
as in the ACT. In Section III we obtain the conditions fulfilled by the
solution: types of spinors and flux quantization. Section III, the origin and
conditions whether the quantum Hall effect appears from the "in plane"
magnetic field are explicitly shown. Magnetic field parallel and the quantum
ring as an application (e.g. spin-filter) is the focus of Section V. The
remanent Section are devoted to discuss, give some concluding remarks and perspectives.

\section{Magnetic field "in plane"}

Now the magnetic field $B,$ in contrast to the ACT\ configuration described
before, is parallel to the plane defined by $x,y$ axis (usually denominated:
"$B$ in plane") where the particle lives. Explicitly the Dirac equation with
the magnetic field parallel takes the following form:
\begin{equation}
\left[  \sigma_{B}\partial_{B}+\sigma_{\perp}\left(  \partial_{\perp
}-ieA_{\perp}\right)  -ie\sigma_{z}A_{z}\right]  \varphi=0 \tag{1}%
\end{equation}
here, the subscripts $B,$ $\perp$ and $z$ denote the direction of the $B$
field in the plane, the direction of the component of the potential vector in
the plane (obviously, perpendicular to the $B$ direction) and the direction of
component of the potential vector coincident with the $z$ axis, respectively.

Defining $\omega$ the angle of the magnetic field with respect the $x$ axis in
the plane $x-y$, the transformation (B) takes in this case, the following
general form.%
\begin{equation}
\psi=e^{i\left(  \alpha\sigma_{x}+\beta\sigma_{y}\right)  }\varphi
=e^{ie\phi\cdot\sigma_{B}}\varphi\tag{2}%
\end{equation}
with
\begin{align}
\alpha &  =\lambda\cos\omega
,\ \ \ \ \ \ \ \ \ \ \ \ \ \ \ \ \ \ \ \ \ \ \ \ \beta=\lambda\sin
\omega\tag{3}\\
\left\vert \phi\right\vert ^{2}  &  =\lambda^{2}\left(  \cos^{2}\omega
+\sin^{2}\omega\right)  =\lambda^{2}\Rightarrow\left\vert \phi\right\vert
=\pm\left\vert \lambda\right\vert \tag{4}%
\end{align}
e.g the projection of the field $\phi$. Equation (1) explicitly written having
account (2), is%
\begin{equation}
\left[  \sigma_{x}\partial_{x}+\sigma_{y}\partial_{y}-ieA_{\perp}\left(
\sigma_{x}\sin^{2}\omega+\sigma_{y}\cos^{2}\omega\right)  -ie\sigma_{z}%
A_{z}\right]  \varphi=0 \tag{5}%
\end{equation}
It is easily seen that, when\ $\omega=0$ $\ B$ coincides with $x-axis$ and
when $\omega=\pi/2$ , $B$ coincides with the $y-axis$) Also the Lie algebraic
relation holds
\begin{equation}
\sigma_{B}\sigma_{\perp}=\left(  \cos\omega\sigma_{x}+\sin\omega\sigma
_{y}\right)  \left(  -\sin\omega\sigma_{x}+\cos\omega\sigma_{y}\right)
=i\sigma_{z} \tag{6}%
\end{equation}
as expected.

Operating analogically as in the ACT\ configuration but having into account
the new transformation and the physical situation, we obtain the conditions
where the magnetic field can be eliminated. Precisely using expression (2) in
(1) we obtain explicitly the following non trivial conditions in order to
remove the magnetic field
\begin{equation}
-\partial_{\perp}\phi=iA_{z}\ \ \ \ \ \ \ \ \ \ \ \partial_{B}\phi=-A_{\perp
}\sigma_{\perp} \tag{7}%
\end{equation}
\ \ \ \ \ \ \ \ \ \ The first equation is precisely as in the ACT\ case but
for the second one the interpretation is more involved and suggest, in
principle, a complex structure for the field $\phi$ in a doublet form. Knowing
that the doublet can be written as%
\begin{equation}
\phi\equiv\left(
\begin{array}
[c]{c}%
\phi_{1}\\
\phi_{2}%
\end{array}
\right)  \tag{8}%
\end{equation}
the previous expressions take the following explicit form%
\begin{equation}
-\partial_{\perp}\phi_{1}=-\partial_{\perp}\phi_{2}=iA_{z}\text{ and
\ \ \ \ }\partial_{B}\phi_{1}=-\text{\ }\partial_{B}\phi_{2}=iA_{\perp}
\tag{9}%
\end{equation}
Notice that above condition, in general, suggest the introduction of 2 real
functions $u$ and$v$ as
\begin{equation}
\phi\equiv\left(
\begin{array}
[c]{c}%
\phi_{1}\\
\phi_{2}%
\end{array}
\right)  =\left(
\begin{array}
[c]{c}%
\phi_{1}\\
\phi_{1}^{\ast}%
\end{array}
\right)  =\left(
\begin{array}
[c]{c}%
u\left(  x_{\perp}\right)  +iv\left(  x_{B}\right) \\
\left(  u\left(  x_{\perp}\right)  +iv\left(  x_{B}\right)  \right)  ^{\ast}%
\end{array}
\right)  \tag{9}%
\end{equation}
in such a manner that the the conditions to remove the magnetic field are
automatically fulfiled as%
\begin{equation}
-\partial_{\perp}\phi=iA_{z}\text{ and \ \ \ \ }\partial_{B}\phi=A_{\perp}
\tag{10}%
\end{equation}
Equation (9b) is nothing more that the Majorana condition over $\phi$ that
appear as consequence of the existence of a parallel magnetic field.

\subsection{Structure of the magnetic field: conditions over $A$ and $\phi$}

The magnetic field that can be effectively generated $(B=\nabla\wedge A)$ from
the vector potential components of our problem, namely $A_{z}$ and $A_{\perp
}.$

The "in plane" magnetic field is consequently
\begin{equation}
B_{B}=(\partial_{\perp}A_{z}-\text{\ }\partial_{z}A_{\perp})\text{\ \ }
\tag{11}%
\end{equation}

where the simplest possibility was took: $A\neq A\left(  x_{B}\right)  $ e.g.
the vector potential does not depends on the direction of the magnetic field,
only on the plane defined by \ $x_{\perp}$ and $x_{z}$(consistly as for the
magnetic field in the ACT\ configuration). From (10) follows
\begin{equation}
B=i\partial_{\perp}^{2}\phi=\frac{\Phi}{x_{\perp}} \tag{12}%
\end{equation}
where the total transversal flux to the plane per unit of longitude was used
\ Then, $\phi$ is immediately obtained%
\begin{equation}
\phi\cdot\sigma_{B}=-i\left(  \Phi\sigma_{B}\right)  x_{\perp}\left[
\ln\left\vert \frac{x_{\perp}}{l_{0}}\right\vert -\frac{C-x_{\perp}}{x_{\perp
}}\right]  \tag{13}%
\end{equation}
Putting the arbitrary constant $C=0$ for simplicity, the behaviour of the
exponential function in (2) for $\varphi$ determined
\begin{equation}
e^{-ie\phi\cdot\sigma_{B}}=\left\vert \frac{l_{0}}{x_{\perp}}\right\vert
^{\frac{e\Phi}{l_{0}}\sigma_{B}x_{\perp}}e^{-\frac{e\Phi}{l_{0}}\sigma
_{B}x_{\perp}} \tag{14}%
\end{equation}
with $l_{0}$ some real constant with units of lengh (its physical meaning will
be anlyzed later). As in the ACT\ case, the following condition must be
fulfiled in order that $\varphi$ be normalizable and square integrable
\begin{equation}
\Phi s_{B}\geqslant0 \tag{15}%
\end{equation}
($s_{B}$ is the spin in the B direction) due%
\begin{equation}
\varphi=e^{-ie\phi\cdot\sigma_{B}}\psi\left(  s,w\right)  \tag{16}%
\end{equation}
In the above expression, the function $\psi$ depends on the spin and some
complex variable $w$ to determine from the simple Dirac-Weyl equation obtained
after the procedure of explicit elimination of the magnetic field.

\subsection{Majorana, Dirac-Weyl states and discrete coordinates: conditions
over $\psi\left(  s,z\right)  $}

the simple Dirac-Weyl equation obtained after "removing the magnetic field "is%
\begin{equation}
\left(  e^{-ie\phi\cdot\sigma_{B}}\sigma_{B}\partial_{B}+e^{ie\phi\cdot
\sigma_{B}}\sigma_{\perp}\partial_{\perp}\right)  \psi\left(  s,z\right)  =0
\tag{17}%
\end{equation}
to solve the equation a quantization should be imposed on the flow (strictly
on the product $\phi\cdot\sigma_{B}$). This fact will induce an automatic
discretization over the in plane transverse coordinate $x_{\perp}.$
\begin{equation}
\phi\cdot\sigma_{B}=n\pi,\text{ \ \ \ \ \ \ }n=0,1,2,,,, \tag{18}%
\end{equation}
If above condition holds, we obtain%
\begin{equation}
\left(  \sigma_{B}\partial_{B}+\sigma_{\perp}\partial_{\perp}\right)
\psi\left(  s,z\right)  =0 \tag{19}%
\end{equation}
This expression is very important: this is a simple 2 dimensional Dirac
equation \textit{without} $A_{\mu}.$ The particular phase introduced as ansatz
plus a quantization condition indicate that the effect of the magnetic field
(due the potential vector )can be removed.

\subsection{Analysis of the solution}

Looking the specific form of the above equations, there are two possibilities
over the spin behaviour of $\psi$

i) $\sigma_{B}\psi\left(  s,z\right)  =s\psi\left(  s,z\right)  $ (eigenspinor
of $\sigma_{B})$

this case is compatible with the assumption that the state is eigenvector of
the spin in the magnetic field direction. The Dirac equation is reduced to
\begin{equation}
\left(  \partial_{B}+\frac{i\mathbb{C}}{s}\partial_{\perp}\right)  \psi\left(
s,z\right)  =0 \tag{20}%
\end{equation}
with $\mathbb{C}$ the charge conjugation operator. Then, $\psi\left(
s,z\right)  $ , and for instance $\varphi\left(  s,z\right)  $must fulfill the
Majorana condition:%
\begin{equation}
\mathbb{C}\varphi\left(  s,z\right)  =\pm c\varphi\left(  s,z\right)  \tag{21}%
\end{equation}

Similarly as in the AC case, $\psi\left(  s,z\right)  $ is an entire function
of $z=x_{B}+\frac{ic}{s}x_{\perp}$ but the states solution is of Majorana type.

ii) $\sigma_{z}\psi\left(  s,z\right)  =s\psi\left(  s,z\right)  $
(eigenspinor of $\sigma_{Z})$in this case the spin remains as in the ACT
situation (e.g. in the z direction). Now the Dirac equation is reduced to
\begin{equation}
\left(  \partial_{B}+is\partial_{\perp}\right)  \psi\left(  s,z\right)  =0
\tag{22}%
\end{equation}
Similarly as in the AC case, $\psi\left(  s,z\right)  $ is an entire function
of $z=x_{B}+isx_{\perp},$ and the state solution is Dirac-Weyl.

The specific form of the equation (20) shows that the result is not casuality:
the states are Majorana. The inclusion of the charge conjugation operator
$\mathbb{C}$ due the symmetry of the physical scenario, enforces obviously,
the Majorana condition over the states solution.

\section{Quantum Hall effect and the "in plane" magnetic field}

Is not difficult to see that, if the plane where the charges are moving is
finite an "in plane" current transversal to the magnetic field B must appear
(e.g in the .$x_{\perp}$ direction$)$ This current will be quantized due the
condition (18). This condition explicitly can be written
\begin{equation}
\phi\cdot\sigma_{B}=\left(  \Phi\sigma_{z}\right)  \widetilde{x}_{\perp
}\left[  \ln\left\vert \frac{x_{\perp}}{l_{0}}\right\vert -1\right]
=n\pi,\text{ \ \ \ \ \ \ }n=0,1,2,,,, \tag{23}%
\end{equation}
where$\widetilde{x}_{\perp}=$ $\sigma_{\perp}x_{\perp}$ is a new matrix
valuated coordinate that its meaning will be analyzed later.

The explicit formula for the Hall current coming from the expression for the
surface current
\begin{equation}
n\times B=K_{surface} \tag{24}%
\end{equation}
($n$: unitary vector normal to the interface surface) this current is
obviously perpendicular to the "in plane" magnetic field (e.g.$x_{\perp}$
direction). Due the quantization condition, the "emergent" Hall current also
is quantized leading the QHE
\begin{equation}
\frac{\Phi}{x_{\perp}}\overset{v}{x}_{\perp}=\frac{2\pi N\hbar c}{ex_{\perp}%
}\overset{v}{x}_{\perp}=K_{surface} \tag{25}%
\end{equation}
where is the versor in the $x_{\perp}$ direction.

\subsection{Generalized momentum operators and Majorana conditions}

The interpretation of the non standard Dirac equation
\begin{equation}
\left[  \sigma_{B}\partial_{B}+\sigma_{\perp}\left(  \partial_{\perp
}-ieA_{\perp}\right)  -ie\sigma_{z}A_{z}\right]  \varphi=0 \tag{26}%
\end{equation}
can be elucidated rewritten it as
\begin{align}
\left[  \sigma_{B}\underset{\widetilde{\Pi}_{B}}{\underbrace{\left(
\partial_{B}-ie\sigma_{B}\sigma_{z}A_{z}\right)  }}+\sigma_{\perp}%
\underset{\Pi_{\perp}}{\underbrace{\left(  \partial_{\perp}-ieA_{\perp
}\right)  }}\right]  \varphi &  =0\Rightarrow\tag{27}\\
\left[  \sigma_{B}\widetilde{\Pi}_{B}+\sigma_{\perp}\Pi_{\perp}\right]
\varphi &  =0\nonumber
\end{align}
\bigskip then, the question that immediately appears is: who is the operator
$\widetilde{\Pi}_{B}$? The answer is obvious, using the algebra (6)$\sigma
_{B}\sigma_{z}=i$ $\sigma_{\perp}$and the definition of the charge conjugation
operator as function of the sigma matrices, is easy to see that%
\begin{equation}
\left(  \partial_{B}-ie\sigma_{B}\sigma_{z}A_{z}\right)  =\left(  \partial
_{B}+ie\mathbb{C}A_{z}\right)  \tag{28}%
\end{equation}
As in ordinary non abelian gauge theories, the operator $\widetilde{\Pi}_{B}$
seems as equipped with a \textbf{non abelian} vector potential $\widetilde
{A_{B}}\equiv-\mathbb{C}A_{z}.$ This conceptual interpretation will be
utilized in the next section for the analisis of the quantum ring.

\section{Magnetic field parallel and the quantum ring}

Was recently pointed out[6], that the Rashba and Dresselhaus spin-orbit
interactions in two dimensions can be regarded as a non-Abelian gauge field
reminiscent of the standard Yang-Mills in QFT [13]. The explanation given in
such references is that the physical field generated by the gauge field brings
to the electron wave function a spin-dependent phase. This phase generally is
called the Aharonov-Casher phase. In ref. [6] the authors showed that applying
on an AB ring this non-Abelian field together with the usual vector potential,
is certainly possible make the interference condition completely destructive
for one component of the spin while completely constructive for the other
component of the spin over the entire energy range. This enables us to
construct a perfect spin filter. However in [6] the magnetic field was\textbf{
perpendicular} to the plane of the ring. \ Now we will proceed analogously but
considering the \textbf{in plane} magnetic field to see the physical
consequences over the physical states and over the spin control. In order to
perform the analysis and to compare with the case described in [6], the same
method and definitions will be used remiting to the reader to ref. [6] for
more details.

The general Dirac equation whith the magnetic field parallel ("in plane") in
cylindrical coordinates takes the form%

\begin{gather}
\left[  \sigma_{\rho}\partial_{\rho}+\frac{1}{\rho}\sigma_{\varphi}%
\partial_{\varphi}-ie\underset{\propto\sigma_{\perp}A_{\perp}}{\underbrace
{\left(  -\sigma_{\rho}A_{\rho}\sin(\omega-\varphi)+\sigma_{\varphi}%
A_{\varphi}\cos(\omega-\varphi)\right)  }}-ie\sigma_{z}A_{z}\right]
\widehat{\varphi}=\tag{29}\\
\left[  \sigma_{\rho}\partial_{\rho}+\frac{1}{\rho}\sigma_{\varphi}\left(
\partial_{\varphi}-ie\sigma_{\rho}\underset{2iA_{z}}{\underbrace
{\widetilde{A_{z}}}}\right)  -ie\left(  -\sigma_{\rho}A_{\rho}\sin
(\omega-\varphi)+\sigma_{\varphi}A_{\varphi}\cos(\omega-\varphi)\right)
\right]  \widehat{\varphi}=0 \tag{30}%
\end{gather}
where in the last equation the propierties of the algebra described in the
previous paragraph have been used in order to introduce the "non abelian"
potential. Notice that in our case is not only a trick in sharp contras with
other refences in the literature. The explicit form of the Pauli matrices in
the configuration that we are interested in are%

\begin{align}
\sigma_{\rho}  &  =\sigma_{x}\cos\varphi+\sigma_{y}\sin\varphi\tag{31}\\
\sigma_{\varphi}  &  =\sigma_{y}\cos\varphi-\sigma_{x}\sin\varphi\tag{32}\\
\sigma_{B}  &  =\sigma_{\rho}\cos(\omega-\varphi)+\sigma_{\varphi}\sin
(\omega-\varphi)\tag{33}\\
\sigma_{\perp}  &  =\sigma_{\varphi}\cos(\omega-\varphi)-\sigma_{\rho}%
\sin(\omega-\varphi) \tag{34}%
\end{align}

However,\ $\varphi$ is the angular cilyndrical coordinate, $\omega$ is the
angle of the magnetic field parallel to the plane measured from the axis $x$
(e.g. $\varphi=0)$ and the state is denoted as $\widehat{\varphi}.$

For the ring configuration, and having account the condition that the
potential doesn't depends on the direction of the magnetic field, the Dirac
equation takes this non abelian form%

\begin{equation}
\left[  \frac{1}{\rho}\sigma_{\varphi}\left(  \partial_{\varphi}%
-ie\sigma_{\rho}\widetilde{A_{z}}\right)  \right]  \widehat{\varphi}=0
\tag{35}%
\end{equation}
then, the corresponding second order equation suggests the Hamiltonian for the
magnetic field in plane as%
\begin{equation}
\mathcal{H}_{ring}=\frac{1}{\rho^{2}}\left(  \partial_{\varphi}-ie\sigma
_{\rho}\widetilde{A_{z}}\right)  ^{2} \tag{36}%
\end{equation}
Notice the non-abelian character of the above equation, that was described in
the previous paragraph.

\subsection{Screening of Rashba term and "in plane' magnetic field}

When the Rashba spin-orbit interaction is introduced, the following
Hamiltonian is obtained%
\begin{equation}
\left.  \mathcal{H}_{ring}\right\vert _{B_{n-plane}}=\frac{\hslash^{2}%
}{2m^{\ast}R^{2}}\left(  -i\partial_{\varphi}-\sigma_{\rho}\left(
\underset{potential}{\underbrace{\widetilde{A_{z}}}}+\underset{Rashba}%
{\underbrace{\frac{\theta R}{2}}}\right)  \right)  ^{2} \tag{37}%
\end{equation}
where $\theta\equiv\frac{2m^{\ast}\alpha}{\hslash}$ plays the role of coupling
constant of the Rashba term (the same units as in reference [6]). The main
point is that the vector potential corresponding to the" in plane" magnetic
field (perpendicular to the plane of the ring) is at the same
\textbf{non-abelian} level that the Rashba term.

In the case treated by the authors in [6], the magnetic field is perpendicular
having the Hamiltonian for the ring the following fashion%
\begin{equation}
\left.  \mathcal{H}_{ring}\right\vert _{B_{z}}=\frac{\hslash^{2}}{2m^{\ast
}R^{2}}\left(  -i\partial_{\varphi}-\underset{\frac{e\pi R^{2}B_{z}}{h}%
}{\underbrace{\phi_{B}}}-\left(  \sigma_{\rho}\underset{Rashba}{\underbrace
{\frac{\theta R}{2}}}\right)  \right)  ^{2} \tag{38}%
\end{equation}
where is easily seen that $\phi_{B}$ is not at the same "non abelian" level of
the Rashba term. Consequently, is this the explanation of the screening of the
Rashba interaction by the "in plane" magnetic field.

\subsection{Physical consequences and effects}

We know from Section II that we can select solutions that are eigenfunctions
of $\sigma_{z}.$ Besides this issue, the potential vector $A_{z}$ must come
perpendicularly to the ring plane ($\widehat{z}$ direction) in concordance
with the ACT\ situation. Notice that, in sharp constrast with previous
references, \textit{the effective Hamiltonian arises from the true Dirac
equation\ with minimal coupling}. As we can assume in general that we know the
magnetic field $\left(  B_{pl}\right)  $ in the plane : $e\widetilde{A_{z}%
}\equiv2ieA_{z}=2ieB_{pl}R/h,$ then%
\begin{equation}
\left.  \mathcal{H}_{ring}\right\vert _{B_{n-plane}}=\frac{\hslash^{2}%
}{2m^{\ast}R^{2}}\left[  -i\partial_{\varphi}-\sigma_{\rho}\left(
\frac{\left(  \theta-4eB_{pl}\right)  R}{2}\right)  \right]  ^{2} \tag{39}%
\end{equation}
the interplay between the magnetic field parallel to the plane of the ring and
the Rashba interaction is clearly seen.

Assuming free interaction into the 2 leads, in response to%
\begin{equation}
\mathcal{H}_{lead}=-\frac{\hslash^{2}}{2m^{\ast}}\partial_{x}^{2} \tag{40}%
\end{equation}
we obtain (same units and notation that in ref.[6]) for the ring Hamiltonian
the wave functions%
\begin{equation}
\Psi_{\pm\pm}=e^{i\left(  \pm k_{\varphi}\pm\phi_{T}\right)  \varphi
}e^{-i\beta\sigma_{\varphi}/2}\chi_{\pm} \tag{41}%
\end{equation}
($\chi_{\pm}$ eigenfunctions of $\sigma_{z})$ \ with the eigenvalues%
\begin{equation}
E=\frac{\hslash^{2}k_{\varphi}^{2}}{2m^{\ast}\rho^{2}} \tag{42}%
\end{equation}
where now the total phase is%
\begin{equation}
\phi_{T}=\sqrt{1+\left(  \theta-4eB_{pl}\right)  ^{2}\rho^{2}}-1 \tag{43}%
\end{equation}
and%
\begin{gather}
\beta=\arctan\xi\tag{44}\\
\text{ with \ }\left(  \theta-4eB_{pl}\right)  \equiv\xi\tag{45}%
\end{gather}
Notice the important fact that the total phase$\phi_{T}$ is
\textbf{identically zero} if the following condition holds%
\begin{equation}
\theta=4eB_{pl} \tag{46}%
\end{equation}
As in [6] the first sign of $\Psi_{\pm\pm}$ denotes the sign of the momentum,
and the second one of the spin. In the phase corresponding to Rashba
interaction, an small radius $\rho$ of the ring was considered. \ Following
similar task that in [6] in order to realize a perfect spin filter, the wave
function (41) at $\varphi=2\pi$ is
\begin{equation}
\Psi_{\pm\pm}\left(  2\pi,k_{\varphi}\right)  =e^{\pm2i\pi k_{\varphi}%
}U_{phase}\chi_{\pm}\text{ \ \ \ \ \ ;\ \ \ \ \ \ \ \ \ \ \ }U_{phase}%
=e^{\pm2\pi i\phi_{T}}e^{-i\beta\sigma_{y}/2} \tag{47}%
\end{equation}
realizing the spin filter by adjusting parameters:%
\[
2\pi\phi_{T}=(2n+1)\pi
\]
At this point, two important cases must be considered:

\subsubsection{Case a )}

Considering (43) and small radius $\rho,$ the above condition is translated to%
\begin{align}
\xi\rho &  =\sqrt{n+3/2},\text{ \ \ \ }n\text{ }\in\mathbb{Z}\tag{48}\\
&  =\sqrt{\frac{3}{2}},\sqrt{\frac{5}{2},}\sqrt{\frac{7}{2}}......\nonumber
\end{align}

\subsubsection{Case b)}

Case (a) must be complemented with a condition that only appears as an effect
of the existence of the magnetic field in plane that is%
\[
\xi=0\rightarrow\theta=4eB_{pl}%
\]
Both conditions, realize the perfect spin filter being the second condition
possible \textbf{only} in the case when the magnetic field is "in plane", and
it importance will be more evident in the coefficient transmission
description, as follows.

The eigenvectors of the phase factor $U_{phase}=e^{\pm2\pi i\phi_{T}%
}e^{-i\beta\sigma_{y}/2}$ can be exactly computed,%
\begin{align}
\widetilde{\chi}_{+}  &  =\left(
\begin{array}
[c]{c}%
\frac{\sqrt{\xi^{2}+1}+1}{2}\\
\xi/2
\end{array}
\right) \tag{49}\\
\widetilde{\chi}_{-}  &  =\left(
\begin{array}
[c]{c}%
\xi/2\\
-\frac{\sqrt{\xi^{2}+1}+1}{2}%
\end{array}
\right)  \tag{50}%
\end{align}
Notice that when the critical value $\theta=4eB_{pl}$ holds, then $\xi=0$
consequently the eigenvectors $\widetilde{\chi}_{\pm}$ goes automatically to
$\chi_{\pm}$ (eigenvectors of $\sigma_{z}$) as expected in sharp contrast with
similar results in [6] that they correctness is doubtful. Although there are
several effective manners to compute the transmission coefficients, we
following ref. [6] in order to compare the results with other works involving
the similar devices. We first assume the amplitudes of the left-going and
right-going wave functions separately for the left lead, the portion
$0<\varphi<\pi$ of the ring, the portion$\pi<\varphi<2\pi$ of the ring, and
the right lead. This amounts to sixteen amplitudes in total when we take the
spin degree of freedom into account. The continuation of the wave function at
$\varphi=0$ and$\varphi=\pi$ give eight conditions and the conservation of the
generalized momentum at $\varphi=0$ and $\varphi=0$ give four conditions.
Then, four degrees of freedom finally remain. The S-matrix of the quantum ring
is obtained by expressing the four amplitudes of the out-going waves (the
left-going wave on the left lead and the right-going wave on the right lead
with spin up and down) in terms of the four amplitudes of the incoming waves
(the right-going wave on the left lead and the left-going wave on the right
lead with spin up and down). The off-diagonal $2\times2$ block of the
$4\times4$ S$-$matrix give the transmission coefficients.In our case, the
transmission coefficients are proportional to
\begin{align}
T_{\widetilde{\upuparrows}},T_{\widetilde{\uparrow\downarrow}}  &
\propto\left\vert 1+e^{2\pi i\phi_{T}}\right\vert ^{2}\tag{51}\\
T_{\widetilde{\downarrow\uparrow}},T_{\widetilde{\downdownarrows}}  &
\propto\left\vert 1+e^{-2\pi i\phi_{T}}\right\vert ^{2} \tag{52}%
\end{align}
where $_{\widetilde{\uparrow}}$ and $_{\widetilde{\downarrow}}$ denote
respectively the spin up $\left(  49\right)  $ and spin down $\left(
50\right)  $ diagonalizing the phase factor $U_{phase}=e^{\pm2\pi i\phi_{T}%
}e^{-i\beta\sigma_{y}/2}.$

Summarizing: in the case a) evidently $T_{\widetilde{\upuparrows}%
},T_{\widetilde{\uparrow\downarrow}},T_{\widetilde{\downarrow\uparrow}%
},T_{\widetilde{\downdownarrows}}=0,$ and in the \ case b) the transmission
coeficients are constant, realizing togrther the perfect spin filter [6,7].

\section{Concluding remarks and outlook}

In this letter the 2-dimensional charge transport with parallel (in plane)
magnetic field was considered. The starting point was reminiscent as the
described in the Aharonov and Casher therorems but with the magnetic field
parallel to the plane of charge transport. In this first paper, several
important results were found of which we can conclude enumerating the
following issues:

i) the specific form of the arising equation enforce the respective field
solution to fulfil the Majorana condition;

ii) when any physical system is represented by this equation the rise of
fields with Majorana type behaviour is immediately explained and predicted;

iii) there exists a quantized particular phase that removes the action of the
vector potential and this produces interesting effects being the Quantum Hall
Effect (QHE)\ one of them that is straighforwardly explained;

iv) the interpretation of Dirac equation with a non abelian electromagnetic
field appears in consequence of the conditions imposed on the fields, not as
an added "by hand";

Also the quantum ring was treated as example where these new effects must appear.

As was mentioned in some references, the term non-abelian was introduced "by
hand" in order to reproduce the effects of the interaction of type RD. In our
case, the "non-abelian" term appears due to the presence of the field parallel
to the transport plane of the charges. Therefore, and as we saw in the problem
of the ring in Section 4, there is competition or screening between the
effects produced by the interaction RD and from the parallel magnetic field.
This competition brings two important consequences, namely:

1) new spin filter effects (different in escence to [6]) , as we have analyzed
from Section 4

2) measurable effects of screening that could give a clear explanation of the
new effects observed in planar nanostructures described in references [9].

In the second part of this letter [10], the relationship between the hidden
symmetries of the particular physical systems described by this equation and
the(e.g. nanostructures, composite particle states etc.)will be discussed and
elucidated with a clear explanation about Majorana, zero modes and supersymmetry.

\section{Acknowledgements}

We are very grateful to the CONICET (Argentina) and the Bogoliubov Laboratory
of Theoretical Physics (BLTP-JINR) for their hospitality and finnantial
support and to the professor R.G. Nazmitdinov for introduce me in the subject
of the transport of charged particles in nanostructures.

\section{References}

[1] Y. Aharonov and A. Casher, Phys. Rev. A\textbf{19}, 2461 (1979)

[2] \textit{Geometric Phases in Physics}, edited by A. Shapere and F.
Wilczek,World Scientific, Singapore, 1989.

[3] M. V. Berry, Proc. R. Soc. London Ser. A \textbf{392}, 45, (1984)

[4] Y. Aharonov and J. Anandan, Phys. Rev. Lett.\textbf{ 58}, 1593, (1987)

[5] Y. Aharonov and A. Casher, Phys. Rev. Lett. \textbf{53}, 319, (1984)

[6] N. Hatano et al, Phys.Rev.A\textbf{75}: 032107, (2007)

[7] T.C. Cheng et al., Phys. Rev B \textbf{76}, 214423, (2011)

[8] Diego Julio Cirilo-Lombardo, Physics Letters B 661, 186-191 (2008);
Foundations of Physics 37: 919-950 (2007); Found Phys (2009) 39: 373--396; The
European Physical Journal C - Particles and Fields, 2012, Volume 72, Number 7,
2079 ; Diego Julio Cirilo-Lombardo with V.I. Afonso, Phys.Lett. A376 (2012)
3599-3603 ; Diego Julio Cirilo-Lombardo and Thiago Prudencio,
Int.J.Geom.Meth.Mod.Phys. 11 (2014) 1450067.

[9] Manuel Valin-Rodriguez and Rashid G. Nazmitdinov, Phys. Rev. B 73, 235306 (2006)

[10] Diego Julio Cirilo-Lombardo, Charge dynamics and "in plane" magnetic
field I, in progress.

[11] E. Majorana, Nuovo Cimento 14 171 (1937).

[12] Stevan Nadj-Perge et al. Science 31 October 2014: Vol. 346 no. 6209 pp. 602-607

[13] T.Fujita et al., J. Appl. Phys. \textbf{110, }121301 (2011)

[14] Marcel Franz, Nature Nanotechnology 8, 149--152 (2013), arXiv:1302.3641,

\end{document}